\documentclass[a4paper,fleqn]{cas-dc}
\usepackage{float}
\usepackage{amsthm}
\usepackage[numbers]{natbib}
\usepackage{amsmath}
\usepackage{todonotes}[inline]
\newtheorem{remark}{Remark}
\usepackage{graphicx}
\usepackage{diagbox} 
\def\tsc#1{\csdef{#1}{\textsc{\lowercase{#1}}\xspace}}
\tsc{WGM}
\tsc{QE}
\tsc{EP}
\tsc{PMS}
\tsc{BEC}
\tsc{DE}
\newtheorem{lemma}{Lemma}

\begin{document}
	\let\WriteBookmarks\relax
	\def\floatpagepagefraction{1}
	\def\textpagefraction{.001}
	\shorttitle{ }
	\shortauthors{Attar et al.}
	\title [mode = title]{Efficient Approximation of the Wigner Kernel in Phase-Space Quantum Mechanics}                      
	
	\tnotetext[1]{This document is the result of the research project funded by Ericsson.}
	
	
	\author[1]{Mehran Attar\corref{cor1}}
	\ead{mehran.attar@etsmtl.ca}
	
	\author[1]{Bassant Selim}
	\ead{bassant.selim@etsmtl.ca}
	
	\author[2]{Jean Michel Sellier}
	\ead{jean.michel.sellier@ericsson.com}
	
	\cortext[cor1]{Corresponding author: mehran.attar@etsmtl.ca}
	
	\affiliation[1]{organization={École de technologie supérieure (ÉTS)},
		city={Montreal},
		state={Quebec},
		country={Canada}}
	
	\affiliation[2]{organization={BCSS AT, Ericsson},
		city={Montreal},
		state={Quebec},
		country={Canada}}
	%
	%
	%
	%
	%
	%
	%
	%
	%
	%
	
	\credit{Data curation, Writing - Original draft preparation}
	
	
	
	
	
	%
	\begin{abstract}
		The Signed Particle Formulation provides a particle-based interpretation of quantum mechanics in phase space, where quantum dynamics are represented through the creation and evolution of signed particles. A central computational challenge in this framework is the evaluation of the Wigner kernel, which generally involves highly oscillatory integrals and can become computationally demanding in time-dependent simulations. This paper proposes an analytical approximation of the Wigner kernel for one-dimensional single body quantum systems by exploiting a series-based representation of the potential function. The resulting expression provides an efficient way to approximate the Wigner kernel and the associated Gamma function, which governs the particle-generation process in the Signed Particle Formulation framework. The proposed approximation is evaluated for several Gaussian-based potential profiles, including single, double, triple, and quadruple Gaussian potentials. Numerical comparisons between the approximated and directly computed Wigner kernels and Gamma functions show that the proposed method captures the main behavior of the exact quantities while significantly reducing the computational cost. These results indicate that the proposed approximation can serve as an efficient computational component for scalable Signed Particle Formulation based quantum simulations.
	\end{abstract}
	
	
	

	\maketitle
	\section{Introduction}
	
	The need for quantum theory arose from experimental observations of extremely small particles, like electrons, that couldn't be explained by classical physics. Concepts such as particle-wave duality and energy quantization, without classical counterparts, puzzled scientists until a set of quantum rules was developed by theoretical physicists, notably E. Schr\"{o}dinger. His introduction of wave functions, along with the Copenhagen interpretation, revolutionized the understanding of quantum systems, making quantum theory the most rigorously tested theory in physics to date.

	Shortly after the Schr\"{o}dinger equation was introduced, alternative formulations of quantum mechanics, such as E. Wigner's approach, emerged. Wigner's formulation bridges classical and quantum physics through quasi-distribution functions, making quantum mechanics more intuitive. However, its mathematical complexity, involving a challenging partial integro-differential equation, limited its application until recent advancements in Monte Carlo techniques made it more accessible \cite{wigner1932quantum, feynman1948space, bohm1952suggested}.

	\subsection{Recent Advances in Quantum Mechanics}
	
	Quantum mechanics emerged out of necessity to explain experiments that classical mechanics could not explain, such as black body radiation, the photoelectric effect, and hydrogen atom spectral lines. These phenomena challenged the validity of Newtonian physics, ultimately leading to the development of what is now called "old quantum mechanics" \cite{einstein1965concerning, stewart1861account}. 
	In the early stages of quantum theory, energy quantization was introduced phenomenologically without rigorous justification. A significant breakthrough came with E. Schr\"{o}dinger's equation (1926) \cite{schr1926quantisierung}, which described quantum systems through complex wave functions, $\psi = \psi(x)$. Schr\"{o}dinger initially interpreted the wave intensity, $\psi^* \psi$, as the electric charge density. Although this provided a distinct image of the electron, it was incorrect, as the linear nature of the equation would cause a charge to spread indefinitely, contradicting experimental evidence that particles are localized in space. While the Schr\"{o}dinger formalism is considered the standard in quantum mechanics, alternative approaches exist. Over time, different, but mathematically equivalent methods have emerged, each with unique strengths and weaknesses. Notable among these are the works of Wigner, Feynman, and Keldysh, who developed approaches based on quasi-distribution functions, path integrals, and non-equilibrium functions, respectively, instead of wave functions. Despite their differences, these methods yield the same predictions as the Schr\"{o}dinger equation. This variety is similar to classical mechanics, where distinct frameworks—Newtonian, Lagrangian, Hamiltonian—can describe a system based on mathematical convenience. 
	
	In particular, the Wigner formalism has been widely applied across various fields, including nuclear physics, chemical reactions, quantum collision theory, and quantum optics \cite{wigner1932quantum}. Recently, it has gained renewed interest for addressing emerging physical problems, such as understanding sub-Planck structures in phase space to explain decoherence. It has also seen applications in nanoelectronics and nanotechnology and has been expanded to handle many-body problems through frameworks such as density functional theory (DFT) and time-dependent ab initio simulations \cite{nedjalkov2004unified, nedjalkov2013wigner}. Its effectiveness lies in enabling fully quantum, multi-dimensional, and time-dependent simulations in phase space \cite{sellier2014many, husimi1940some}.
	
	Building on this phase-space perspective, the Signed Particle Formulation (SPF) provides a particle-based continuation of Wigner's formulation, in which quantum systems are represented by ensembles of positive and negative signed particles evolving through free-flight, creation, and annihilation events. In this sense, SPF can be viewed as a physical and computational interpretation of the Wigner Monte Carlo framework, while preserving its connection to the time-dependent Wigner equation and, equivalently, to the Schrödinger equation \cite{sellier2015signed}.
	
	%
	
	Monte Carlo methods are also used in computational mathematics to approximate solutions by employing random sampling \cite{metropolis1949monte, dimovmonte}. By randomly sampling a variable whose expected value is the target functional, these methods produce statistical estimates, effectively transforming complex problems into calculations of mathematical expectations \cite{kalos2008monte}. They are especially useful for mathematical, physical, and engineering problems where deterministic methods fail, as they don’t require the solution to have any specific regularity and allow accuracy to be controlled in terms of probability error \cite{dimovmonte}. Monte Carlo methods are also highly efficient for handling large-scale problems, such as multi-dimensional integration and large linear systems, and they perform well on parallel processors, making them widely applicable across applied sciences.
	Advances in powerful computing, particularly with parallel machines, have accelerated the development of Monte Carlo methods. Today, Monte Carlo algorithms address a vast array of computational problems, including the solution of the Wigner equation, making it challenging to catalog them all. These methods fall into two primary categories: Monte Carlo simulations and Monte Carlo numerical methods. Monte Carlo simulations model physical processes, acting as tools that replicate the underlying physical, chemical, or biological laws \cite{sellier2015signed, jacoboni1983monte}.
	
	
	
	%
	\subsection{Contribution of The Work}\label{sec:contribution}
	 Despite the increasing adoption of the Wigner Monte Carlo framework and the SPF for simulating quantum systems (see \cite{wang2021solving, jonasson2015dissipative} and references therein), the efficient evaluation of the Wigner kernel remains a central computational challenge. In SPF-based simulations, the Wigner kernel directly determines the associated $\gamma$ function, which governs the stochastic creation of signed particles. However, the direct numerical computation of the Wigner kernel generally involves highly oscillatory integrals, which can impose a significant computational burden and hinder scalability, particularly in time-dependent and large-scale simulations.
		
		To address this limitation, this work builds upon the SPF of quantum mechanics, where quantum systems are represented as ensembles of signed, field-less classical particles~\cite{sellier2015signed}. Within this framework, we propose an analytical approximation of the Wigner kernel for one-dimensional quantum systems. The proposed approximation is derived by approximating the potential-function difference appearing in the Wigner kernel using the leading term of its Maclaurin series expansion. This results in a closed-form expression that avoids costly numerical integration and enables efficient estimation of the associated $\gamma$ function.
		
		The main contribution of this work is therefore twofold. First, it provides a computationally efficient analytical approximation of the Wigner kernel that can be used within SPF-based quantum simulations. Second, it demonstrates, through several Gaussian-based potential profiles, that the proposed approximation can closely reproduce the behavior of the numerically computed $\gamma$ function while significantly reducing the execution time required for kernel evaluation. 
		
		Overall, the proposed method offers a practical approximation tool for improving the computational efficiency of Wigner Monte Carlo and SPF-based quantum simulations while preserving the essential phase-space structure of the underlying quantum dynamics. Since SPF-based simulations rely on a Monte Carlo framework, which is inherently statistical and tolerant to sampling-induced fluctuations, the proposed analytical approximation is sufficiently accurate for capturing the dominant behavior of the Wigner kernel and the associated Gamma function while substantially reducing the computational cost.

	\section{Signed Particle Approach}
	
	In this section, we review the signed particle formulation of quantum mechanics by outlining three defining postulates of the theory \cite{sellier2015signed}:
	
	\begin{itemize}
		\item \textbf{Postulate I:} Physical systems can be represented by virtual Newtonian particles, each with a position \( x \) and momentum \( p \) simultaneously, carrying either a positive or negative sign.
		
		\item \textbf{Postulate II:} A signed particle moving in a potential \( V = V(x) \) acts as a field-less classical point particle. During a time interval \( dt \), it generates a pair of signed particles with a probability \( \gamma(x(t)) \, \).
		\begin{equation}\label{eq:gamma_function}
			\displaystyle
			\gamma(x) = \int_{-\infty}^{+\infty} \mathcal{D}p' \, V_M^+\left(x; p'\right)\equiv \lim_{\Delta p' \to 0^+} \sum_{M=-\infty}^{+\infty} V_M^+\left(x; M \Delta p'\right)
		\end{equation}
		where 
		\begin{equation} \label{eq:wigner_original}
			\displaystyle
			V_W(x;p) = \frac{i}{\pi^d \hbar^{d+1}} \int_{-\infty}^{\infty}  dx' \exp^{-\frac{2i}{\hbar}x'p}[V(x+x') - V(x-x')]
		\end{equation}
		and $V_M^+(x;p)$ is the positive part of the quantity and 
		known as the Wigner kernel (in a d-dimensional space) \cite{wigner1932quantum}. At the moment of creation, if the parent particle has a sign $s$, a position $x$, and a momentum $p$, then the newly created particles will both be located at $x$, will have signs $+s$ and $-s$, and momenta $p + p'$ and $p - p'$, respectively, where $ p'$ is selected randomly based on the normalized probability $\displaystyle\frac{V_W^+ (x; p)}{\gamma(x)} $.

		\item \textbf{Postulate III.} Two particles with opposite sign and same phase-space coordinates $(x,p)$ annihilate.
	\end{itemize}
	
	
	%
	\subsection{Annihilation Technique} \label{sec:annihilation}
	
	It can be shown that the creation of new particle pairs follows an exponential pattern \cite{nedjalkov2004unified}. In the Monte Carlo method described above, particles are indistinguishable and are annihilated if they occupy the same phase-space cell with opposite signs. This approach allows for the elimination of a substantial number of particles during the simulation. The details of this technique are extensively covered in \cite{sellier2014role}, but its main principles are briefly outlined here. By introducing a recording time step, particles in the same phase-space region with opposite signs can be identified and removed, while non-annihilating particles remain in the simulation. This process enables the periodic removal of particles that do not contribute to the calculation of the Wigner function. Essentially, the numerical average of the Wigner quasi-distribution is renormalized through particle annihilation. This aligns with the Markovian nature of the evolution, where the solution at each time step serves as the starting point for the next.

	This method has proven highly effective, particularly in simulating realistic systems involving tens or even hundreds of millions of initial particles. Without it, time-dependent Monte Carlo simulations of the Wigner equation would be virtually impossible \cite{sellier2015signed}.

	\begin{figure}[h!]
		\centering
		\includegraphics[width=1\linewidth]{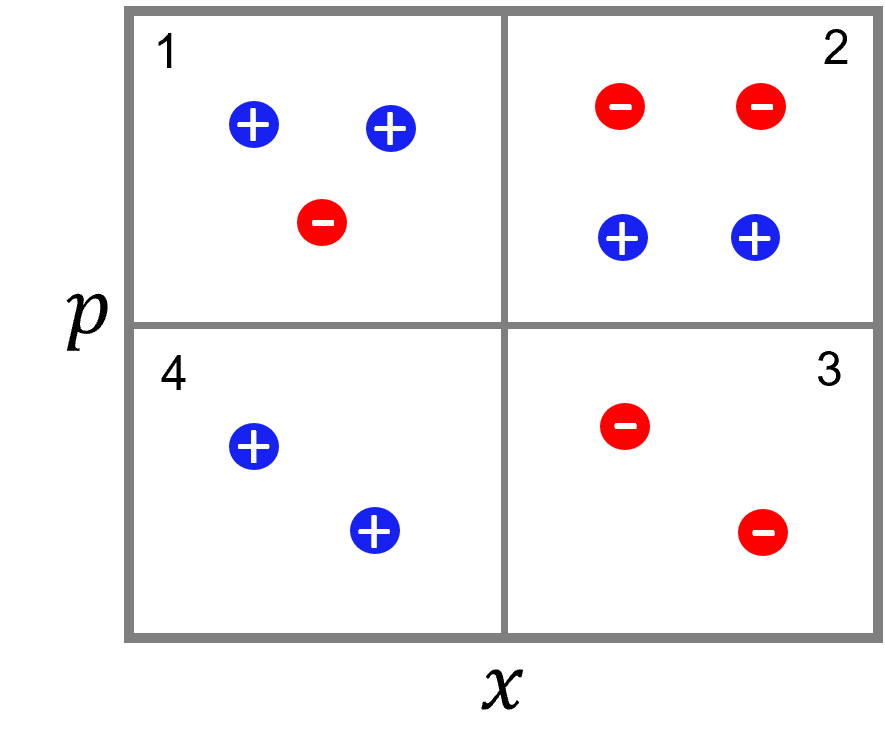}
		\caption{Annihilation of particles: In region (1), one positive particle annihilates a negative particle, instead of a single positive particle.}
		\label{fig:annihilation}
	\end{figure}
	\section{Estimation of The Wigner Kernel} \label{sec:estimation_of_wigner_kernel} 
	
	The Wigner kernel in a one-dimensional, single-body space is defined as follows \cite{sellier2015signed}:
	\begin{equation} \label{eq:wigner_original}
		V_W(x;p) = \frac{i}{\pi \hbar^{2} } \int_{-\infty}^{\infty} e^{-\frac{2i}{\hbar}x'\cdot p} [V(x+x') - V(x-x')] dx' 
	\end{equation}
	%
	%
	Based on the Euler formula, we can rewrite \eqref{eq:wigner_original} as follows: 
	\begin{equation} \label{eq:wigner_after_euler}
		V_W(x;p) = \frac{i}{\pi \hbar^2} \int_{-\infty}^{\infty} (\cos{\theta} - i\sin{\theta}) [V(x+x') - V(x-x')] dx'
	\end{equation}
	where $\theta = \frac{2x'p}{\hbar}$. Moreover, based on the proposed method in \cite{wigner1932quantum}, we can rewrite \eqref{eq:wigner_after_euler} as follows: 
	\begin{equation}\label{eq:applying_quantum_rule}
		V_W(x;p) = \frac{i}{\pi \hbar^2} \int_{-\infty}^{\infty} (- i\sin{\theta}) [V(x+x') - V(x-x')] dx'   
	\end{equation}
	By changing the variable $x' = \frac{pt'}{2m}$ (as proposed in \cite{sellier2015signed}), then equation \eqref{eq:applying_quantum_rule}, can be rewritten in the following format based on $t'$: 
			
	\begin{equation} \label{eq:wigner_based_t_prime}
		\begin{aligned}
			V_W(x;p) =\; & \frac{i}{\pi \hbar^2} \int_{-\infty}^{\infty} (- i\sin{\theta}) \\
			& \times \left[ V\!\left(x+\frac{pt'}{2m}\right) - V\!\left(x-\frac{pt'}{2m}\right) \right] \frac{p}{2m}\, dt'
		\end{aligned}
	\end{equation}
	which can be simplified as follows: 
	\begin{equation}\label{eq:simplified_integ1}
		V_W(x;p) = \frac{1}{\pi \hbar^{2} }\frac{p}{2m} \int_{-\infty}^{\infty} (\sin{\theta}) [V(x+\frac{pt'}{2m}) - V(x-\frac{pt'}{2m})]  dt'
	\end{equation}
	The following Lemma is instrumental to simplify \eqref{eq:simplified_integ1}.
	\begin{lemma}\label{lemma:mclaurent_series}
		The Maclaurin series for 
		$V\left(x+\frac{pt'}{2m}\right)-V\left(x-\frac{pt'}{2m}\right)$
		can be obtained as follows:
		\begin{equation}
			\begin{aligned}
				&V\left(x+\frac{pt'}{2m}\right)
				- V\left(x-\frac{pt'}{2m}\right) \\
				&= \sum_{n=0}^{\infty} \frac{1}{n!}V^{(n)}\big|_{x=0}
				\left(x+\frac{pt'}{2m}\right)^n \\
				&\quad - \sum_{n=0}^{\infty} \frac{1}{n!}V^{(n)}\big|_{x=0}
				\left(x-\frac{pt'}{2m}\right)^n \\
				&= \left(
				V\big|_{x=0}
				+ \frac{1}{1!}V^{(1)}\big|_{x=0}\left(\frac{pt'}{2m}\right)
				+ \frac{1}{2!}V^{(2)}\big|_{x=0}\left(\frac{pt'}{2m}\right)^2
				+ \cdots
				\right) \\
				&\quad - \left(
				V\big|_{x=0}
				+ \frac{1}{1!}V^{(1)}\big|_{x=0}\left(-\frac{pt'}{2m}\right)
				+ \frac{1}{2!}V^{(2)}\big|_{x=0}\left(-\frac{pt'}{2m}\right)^2
				+ \cdots
				\right) \\
				&= \sum_{\substack{n=1 \\ n\ \mathrm{odd}}}^{\infty}
				\frac{2}{n!}V^{(n)}(x)\big|_{x=0}
				\left(\frac{pt'}{2m}\right)^n .
				\qedhere
			\end{aligned}
		\end{equation}\hfill $\blacksquare$
	\end{lemma} 

	Using Lemma~\ref{lemma:mclaurent_series}, the Wigner kernel in \eqref{eq:wigner_based_t_prime} can be rewritten as follows: 
	\begin{equation}\label{eq:wigner_after_mclaurent}
		V_W(x;p) = \frac{p}{2m\pi \hbar^{2} } \int_{-\infty}^{\infty} \!\!\!\!(\sin{\theta}) \!\!\!\! \sum_{n=odd}^{\infty}\!\frac{2}{n!}V^{(n)}(x)|_{x=0}\left(\frac{pt'}{2m} \right)^n\!\! dt'
	\end{equation}
	Considering the first term of the Maclaurin series of the potential function $V$ (i.e., $n=1$), then \eqref{eq:wigner_after_mclaurent} can be rewritten as follows: 
	\begin{equation} \label{eq:mclaurent_n_1}
		V_W(x;p) = \frac{pV'(x)}{2\pi\hbar^2 m}\left( \frac{p}{m}\right) \int_{-\infty}^{\infty} t' \sin\left(\ \frac{p^2 t'}{\hbar m} \right) dt'
	\end{equation}
	To obtain a tractable analytical expression, the integral in \eqref{eq:mclaurent_n_1} is truncated over a finite computational window. Consequently, \eqref{eq:mclaurent_n_1} can be rewritten as follows:
	%
	\begin{equation} \label{eq:final_wigner_after_integral}
		\begin{array}{cc}
			V_W(x;p) = \displaystyle\frac{p^2 V'(x)}{2\pi \hbar^2 m^2}\left[ \frac{-t'\hbar m}{p^2} \cos(\frac{p^2t'}{\hbar m}) + \frac{h^2m^2}{p^4}\sin(\frac{p^2 t'}{\hbar m})  
			\right ]_{-\infty}^{\infty} \\
			\displaystyle\approx \frac{p^2 V'(x)}{2\pi \hbar^2 m^2} \frac{h^2m^2}{p^4}\sin\left(\frac{p^2T}{\hbar m} \right),
		\end{array}
	\end{equation}
	%
	where $T$ shows a \textit{long enough time} \cite{sellier2015signed}. Consequently, the Wigner kernel can be approximated by the following formula:
	\begin{equation} \label{eq:final_result_estimation}
		\begin{array}{cc}
			V_W(x;p) = \displaystyle\frac{V'(x)}{2\pi p^2}\sin\left(\frac{p^2T}{\hbar m} \right) = \\
			\displaystyle\frac{V'(x)}{2\pi (\hbar k)^2}\sin\left(\frac{(\hbar k)^2 (2mx)}{\hbar\hbar k m} \right) = \\
			\displaystyle\frac{V'(x)}{2\pi (\hbar k)^2}\sin\left(2kx \right)
		\end{array}
	\end{equation}
	%
	%
	Assuming $V(x) = -U e^{-\frac{x^2}{2\sigma^2}}$, then the Wigner kernel in \eqref{eq:final_result_estimation} is given as:
	\begin{equation}\label{eq:last_estimation}
		V_W(x;p) = \frac{\hbar U x }{2\pi \sigma^2 p^2}e^{-\frac{x^2}{2\sigma^2}}\sin\left(\frac{p^2T}{\hbar m} \right)
	\end{equation}
	\newline
	Assuming that $t=\frac{2mx}{\hbar k}$ and since $p=\hbar k,$ equation \eqref{eq:last_estimation} can be rewritten as follows \cite{kittel1955solid}: 
	\begin{equation}\label{eq:estimation_with_other_description}
		\begin{array}{cc}
			\displaystyle
			V_W(x;p) = \frac{\hbar U x }{2\pi \sigma^2 \hbar^2 k^2}e^{-\frac{x^2}{2\sigma^2}}\sin\left(\frac{\hbar^2 k^2 T}{\hbar m} \right) = \\
			\displaystyle\frac{U x }{2\pi \sigma^2 \hbar k^2}e^{-\frac{x^2}{2\sigma^2}}\sin\left(\frac{\hbar k^2 T}{ m} \right) = \\
			\displaystyle\frac{U x }{2\pi \sigma^2 \hbar k^2}e^{-\frac{x^2}{2\sigma^2}}\sin\left(2xk \right) 
		\end{array}
	\end{equation}
	%
	\begin{remark}
		The formula derived in Equation \eqref{eq:final_result_estimation} is applicable to a variety of potential functions. Specifically, in Equation \eqref{eq:estimation_with_other_description}, we consider a Gaussian potential and compute an estimation of the Wigner kernel for this function. The numerical results demonstrate that the proposed estimation formula performs exceptionally well for functions involving exponential terms. Furthermore, other potential profiles can be effectively modeled or approximated using alternative methods, such as curve fitting or Fourier series, thereby enhancing the generality and versatility of the approach.
	\end{remark}
	
%
	
	\section{Numerical Results} \label{sec:numerical_results}
	
	In this section, we demonstrate the effectiveness of the proposed approach by comparing the outcomes of the Wigner kernel and Gamma function, computed using the estimated formula, with their true counterparts under the same potential function. To this end, we consider four types of potential functions for the quantum system: (i) a single Gaussian, (ii) a double Gaussian, (iii) a triple Gaussian, and (iV) a quadruple Gaussian. Moreover, here the Wigner kernel is simulated in a discrete phase-space of position $x$ and momentum $p$ with step $\Delta p=\frac{\hbar p}{L_X}$ where $L_X=200$ nm is a free parameter defining the discretization. Please note that, to compute the estimation of the Wigner kernel, we only use the first term of the McLaurian series defined in \eqref{eq:wigner_after_mclaurent}. 
	It is also worth noting that, in Monte Carlo-based simulations, the quantities of interest are obtained through statistical sampling rather than deterministic pointwise evaluation. Therefore, moderate approximation errors in the Wigner kernel may have a limited impact on the overall simulation outcome, provided that the dominant structure of the kernel and the associated Gamma function is preserved.
	\subsection{Single Gaussian Potential}\label{sec:single_Guassian_test}
	
\begin{figure}[h!]
		\centering
		\includegraphics[width=1\linewidth]{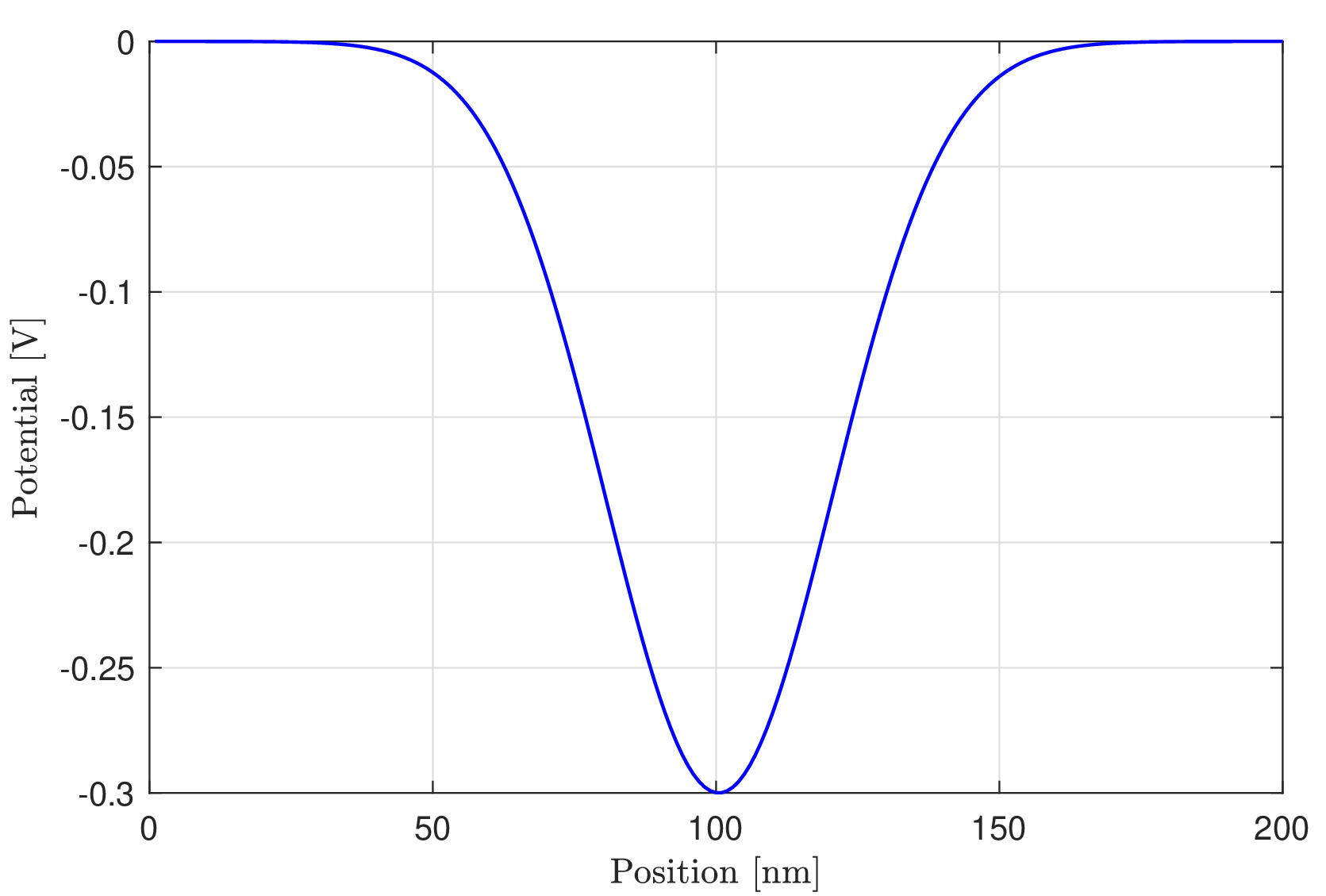}
		\caption{Single Gaussian potential function}
		\label{fig:single_gaussian}
	\end{figure}

\begin{figure}
		\centering
		\includegraphics[width=1\linewidth]{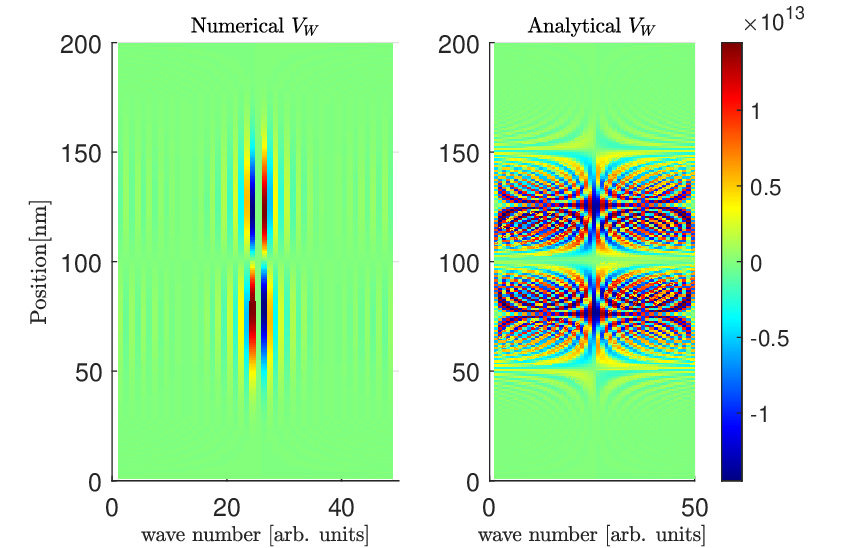}
		\caption{Comparison of actual and estimated Wigner kernel for a single Gaussian potential function.}
		\label{fig:wigner_kernel_single_gaussian}
	\end{figure}

\begin{figure}[h!]
		\centering
		\includegraphics[width=1\linewidth]{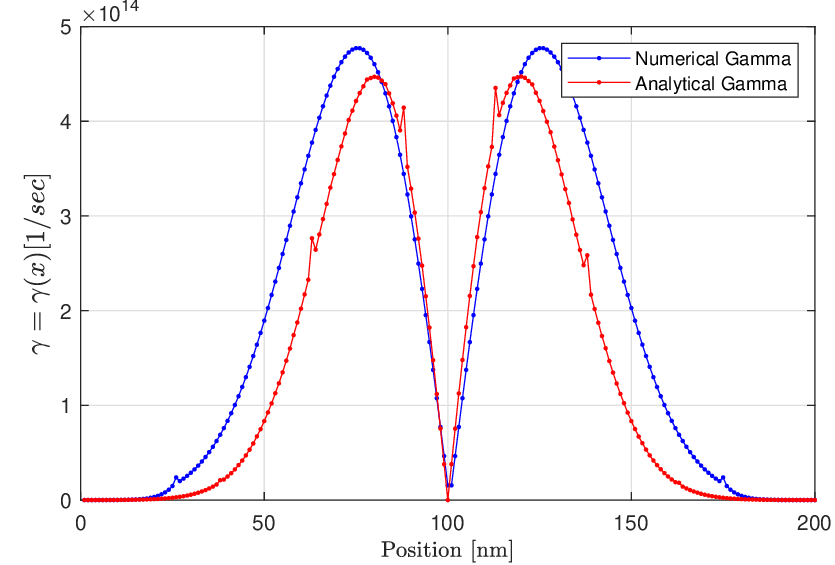}
		\caption{Comparison of Gamma function using the actual and estimated Wigner kernel for a single Gaussian potential function.}
		\label{fig:gamma_single_gaussian}
	\end{figure}

	In this section, we consider a single Gaussian potential function $V(x)$, which is described as
	\begin{equation}\label{eq:signle_gaussian}
		V(x) = V_{max} \cdot \exp\left(-\frac{1}{2} \left(\frac{x - \frac{1}{2} \cdot L_X}{\frac{1}{10} \cdot L_X}\right)^2 \right)
	\end{equation}                          
	where $V_{max}=-0.3$ (see Figure~\ref{fig:single_gaussian}). Now, by plugging \eqref{eq:signle_gaussian} into \eqref{eq:wigner_original} and \eqref{eq:final_result_estimation}, the actual and estimated Wigner kernels are computed, respectively (see Figure \ref{fig:wigner_kernel_single_gaussian}). Then, the actual and estimated Gamma functions can be obtained using \eqref{eq:gamma_function}. Based on Figure \ref{fig:gamma_single_gaussian}, it is observed that that by using the first term of McLaurian series in \eqref{eq:wigner_after_mclaurent}, the estimated Gamma function is close to the numerical one.  
	Finally, it is possible to extract several mathematical properties for the function
	$\gamma(x)$ which would hardly be visible in a purely numerical context. For example, one can easily show that $\gamma(x)=\gamma(-x)$ for all $x$.
	%
	
	%
	%
	

	\subsection{Double Gaussian Potential}
\begin{figure}[h!]
		\centering
		\includegraphics[width=1\linewidth]{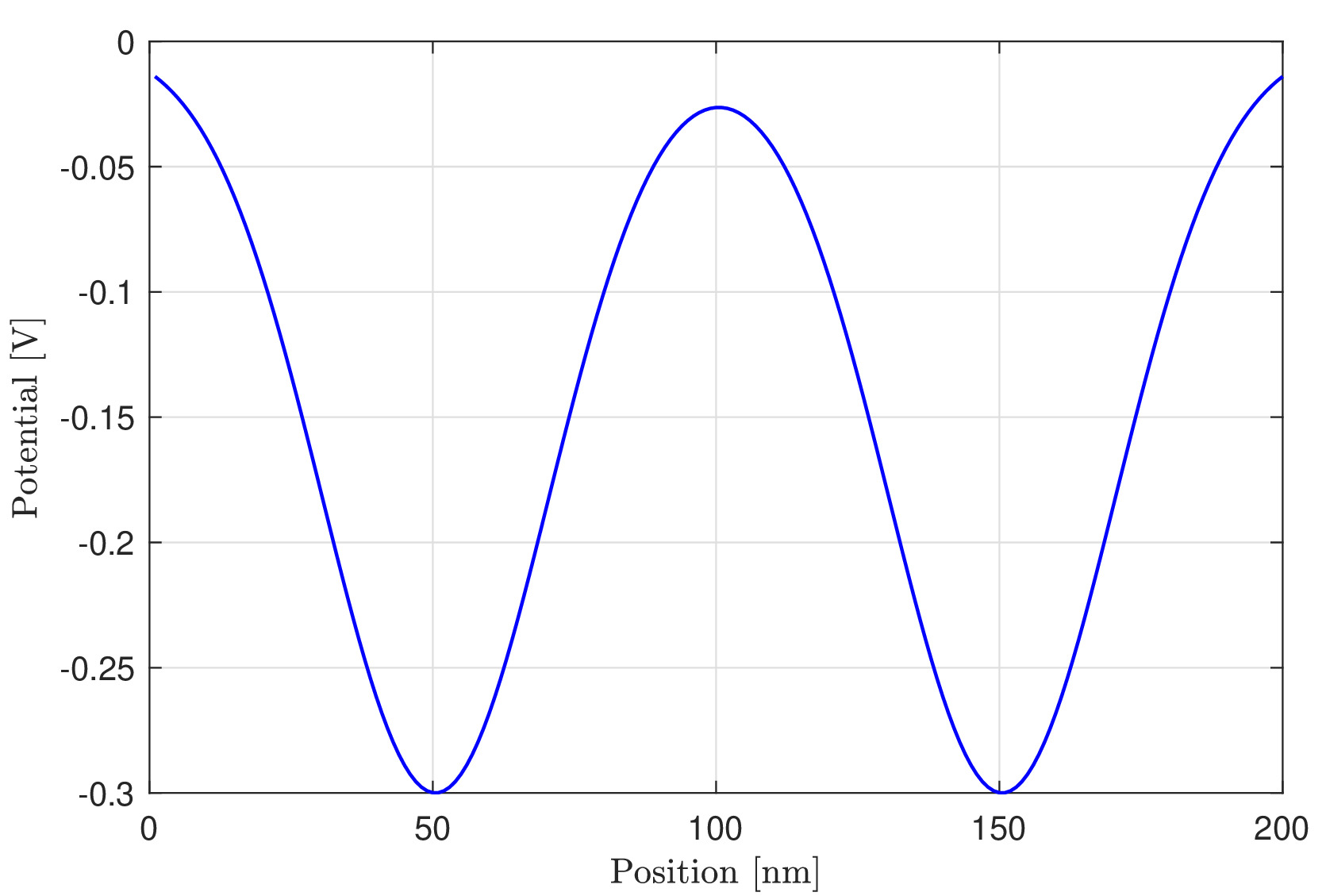}
		\caption{Double Gaussian potential function.}
		\label{fig:double_gaussian}
	\end{figure}

\begin{figure}[h!]
		\centering
		\includegraphics[width=1\linewidth]{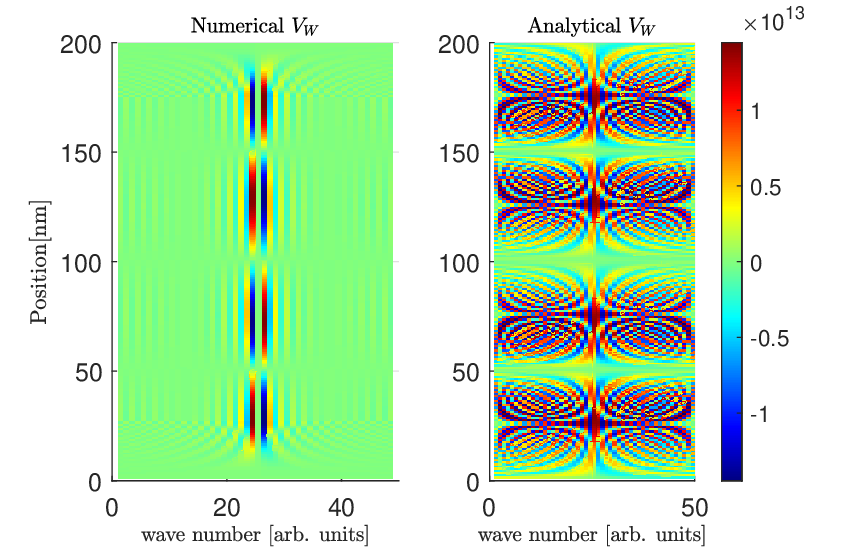}
		\caption{Comparison of actual and estimated Wigner kernel for a double Gaussian potential function.}
		\label{fig:double_gaussian_kernels}
	\end{figure}

\begin{figure}[h!]
		\centering
		\includegraphics[width=1\linewidth]{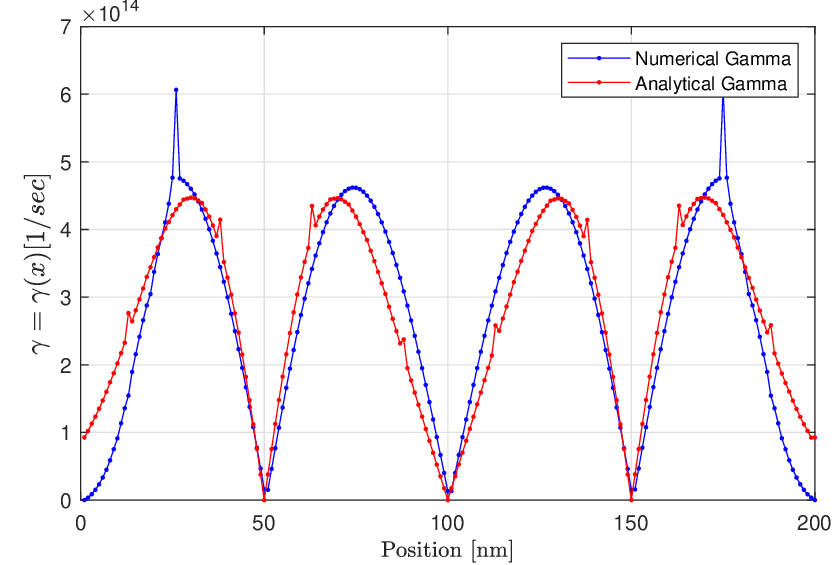}
		\caption{Comparison of the Gamma function using the actual and estimated Wigner kernel for a double Gaussian potential function.}
		\label{fig:double_gaussian_gamma}
	\end{figure}

	In this section, we consider a double Gaussian potential function $V(x)$, which is described as
	\begin{multline} 
		V(x) = V_{max} \cdot \exp\left(-\frac{1}{2} \left(\frac{x - 0.25L_X}{0.1L_X}\right)^2\right)\\ + V_{max} \cdot \exp\left(-\frac{1}{2} \left(\frac{x - 0.75L_X}{0.1L_X}\right)^2\right)
	\end{multline}
	where $V_{max}=-0.3$ (see Figure~\ref{fig:double_gaussian}). Following the method outlined in the previous subsection, we estimate the Wigner kernel and the Gamma function using equation \eqref{eq:final_result_estimation}. Upon comparison with their actual counterparts, it is evident that the estimated Gamma function closely approximates the actual one (see Figure \ref{fig:double_gaussian_kernels} and Figure \ref{fig:double_gaussian_gamma}). This demonstrates the effectiveness of the estimated function derived from \eqref{eq:final_result_estimation} for more complex functions.
	%
	
	%

	\subsection{Triple Gaussian Potential}
	
\begin{figure}[h!]
		\centering
		\includegraphics[width=1\linewidth]{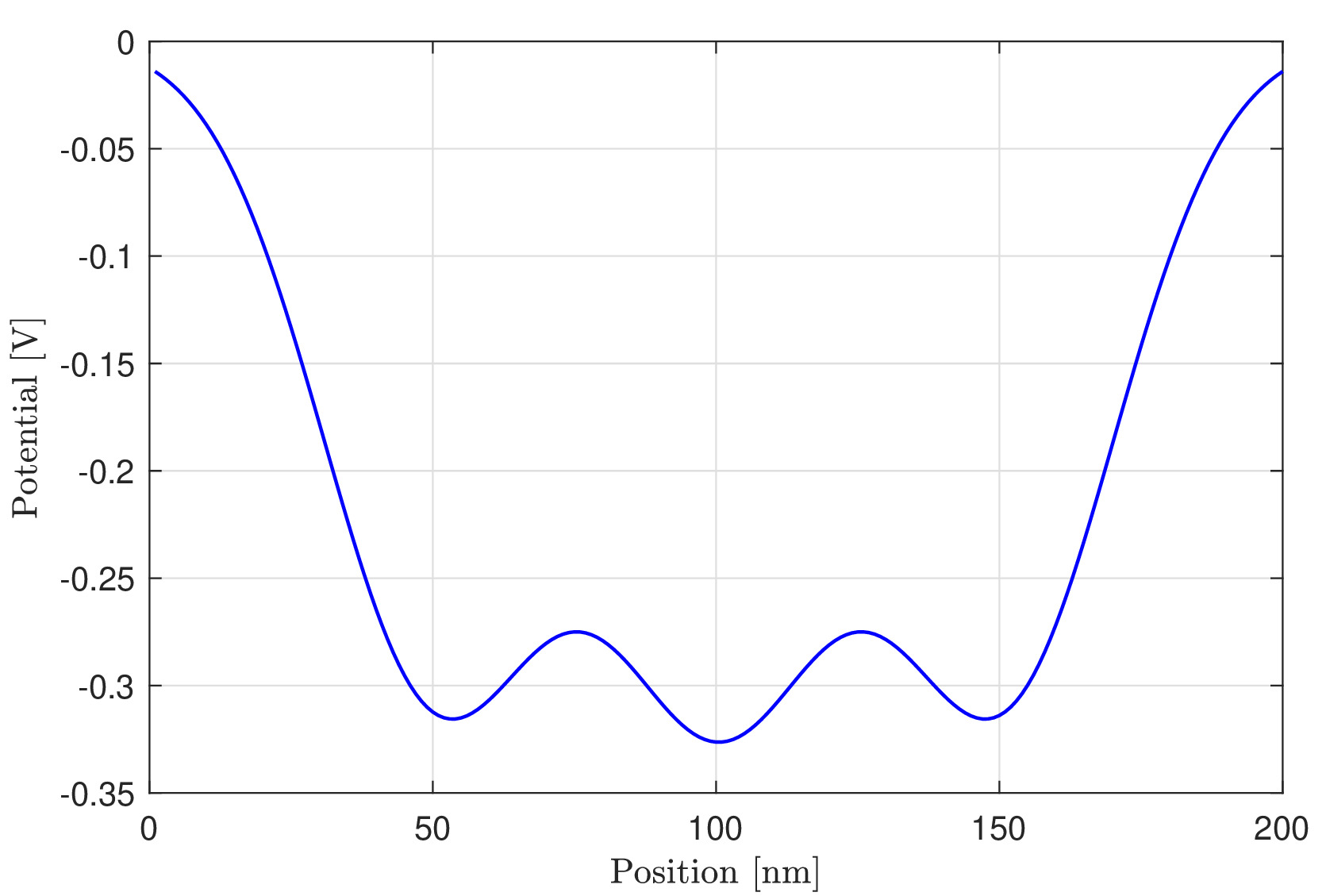}
		\caption{Triple Gaussian potential function.}
		\label{fig:triple_gaussian}
	\end{figure}

\begin{figure}[h!]
		\centering
		\includegraphics[width=1\linewidth]{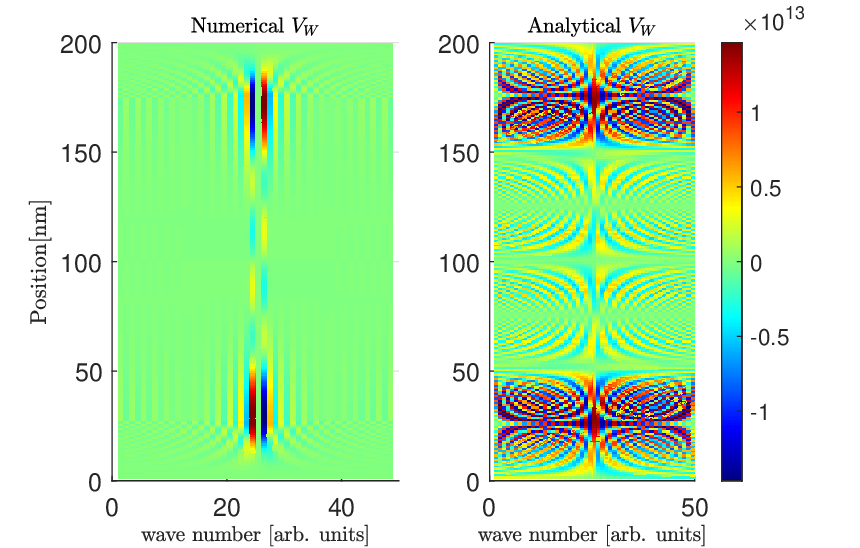}
		\caption{Comparison of actual and estimated Wigner kernel for a triple Gaussian potential function.}
		\label{fig:triple_gaussian_kernel}
	\end{figure}

\begin{figure}[h!]
		\centering
		\includegraphics[width=1\linewidth]{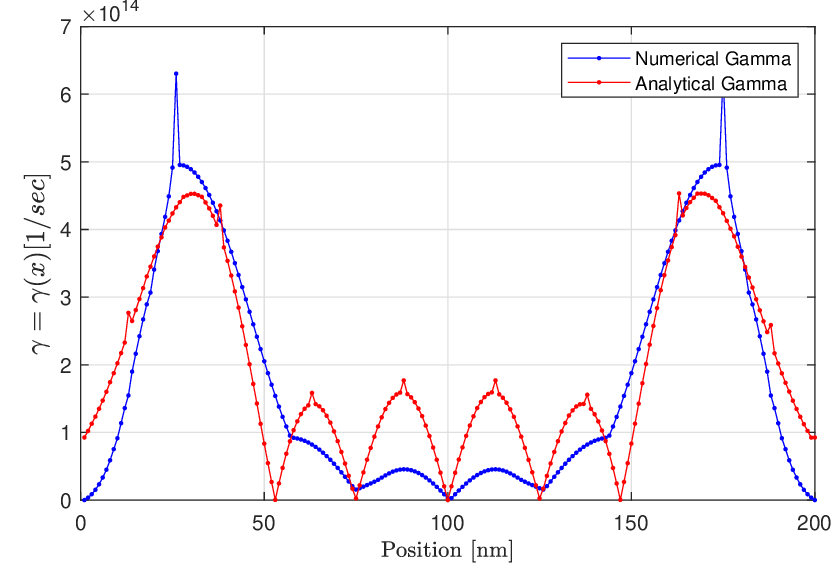}
		\caption{Comparison of Gamma function using the actual and estimated Wigner kernel for a triple Gaussian potential function.}
		\label{fig:triple_gaussian_gamma}
	\end{figure}

	In this section, a triple Gaussian function is considered as a benchmark, which is described as
	\begin{equation}
		\begin{split}
			V(x) =\;& V_{max} \cdot \exp\left(-\frac{1}{2}
			\left(\frac{x - 0.25 L_X}{0.1 L_X}\right)^2 \right) \\
			&+ V_{max} \cdot \exp\left(-\frac{1}{2}
			\left(\frac{x - 0.50 L_X}{0.1 L_X}\right)^2 \right) \\
			&+ V_{max} \cdot \exp\left(-\frac{1}{2}
			\left(\frac{x - 0.75 L_X}{0.1 L_X}\right)^2 \right)
		\end{split}
	\end{equation}
	where $V_{max}=-0.3$ (see Figure~\ref{fig:triple_gaussian}). Using the approach described in subsection \ref{sec:single_Guassian_test}, we estimate the Wigner kernel and the Gamma function based on equation \eqref{eq:final_result_estimation}. A comparison with their true counterparts reveals that the estimated Gamma function closely matches the actual one (refer to Figure \ref{fig:triple_gaussian_kernel} and Figure \ref{fig:triple_gaussian_gamma}). This confirms the effectiveness of the estimation derived from equation \eqref{eq:final_result_estimation}, even for more intricate functions.
	\subsection{Quadruple Gaussian Potential}
	In this section, a quadruple Gaussian function is used as a benchmark, defined as follows:
	
	%
	\begin{multline}
		V(x) = V_{max} \Big[ \exp\left(-\frac{1}{2} \left(\frac{x - 0.125L_X}{0.1L_X}\right)^2\right) \\
		+ \exp\left(-\frac{1}{2} \left(\frac{x - 0.375L_X}{0.1L_X}\right)^2\right) \\
		+ \exp\left(-\frac{1}{2} \left(\frac{x - 0.625L_X}{0.1L_X}\right)^2\right) \\
		+ \exp\left(-\frac{1}{2} \left(\frac{x - 0.875L_X}{0.1L_X}\right)^2\right) \Big].
	\end{multline}

	where $V_{max}=-0.3$ (see Figure~\ref{fig:triple_gaussian}). Using the approach described in the previous section, we estimate the Wigner kernel and the Gamma function based on equation \eqref{eq:final_result_estimation}. A comparison with their true counterparts reveals that the estimated Gamma function closely matches the actual one (refer to Figure \ref{fig:triple_gaussian_kernel} and Figure \ref{fig:triple_gaussian_gamma}). This confirms the effectiveness of the estimation derived from equation \eqref{eq:final_result_estimation}, even for more intricate functions.
	\begin{figure}[h!]
		\centering
		\includegraphics[width=1\linewidth]{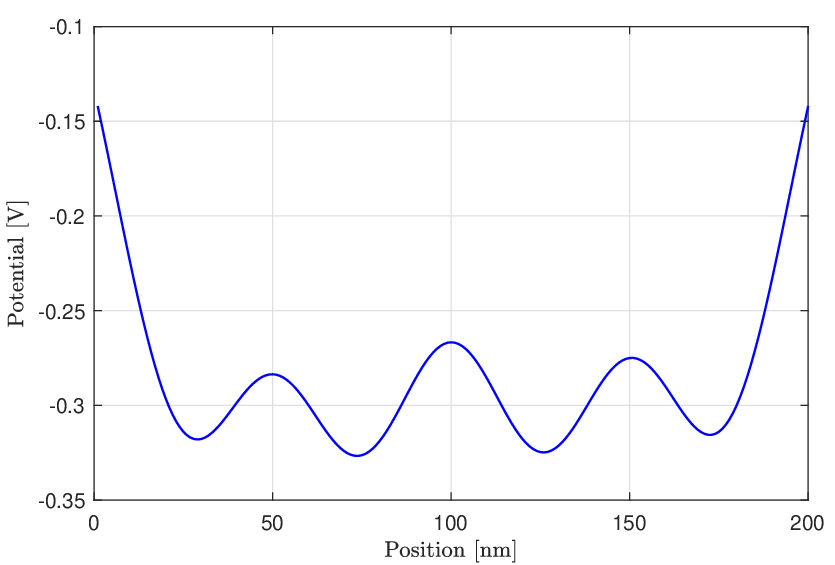}
		\caption{Quadruple Gaussian potential function.}
		\label{fig:quad_gaussian}
	\end{figure}
	\begin{figure}[h!]
		\centering
		\includegraphics[width=1\linewidth]{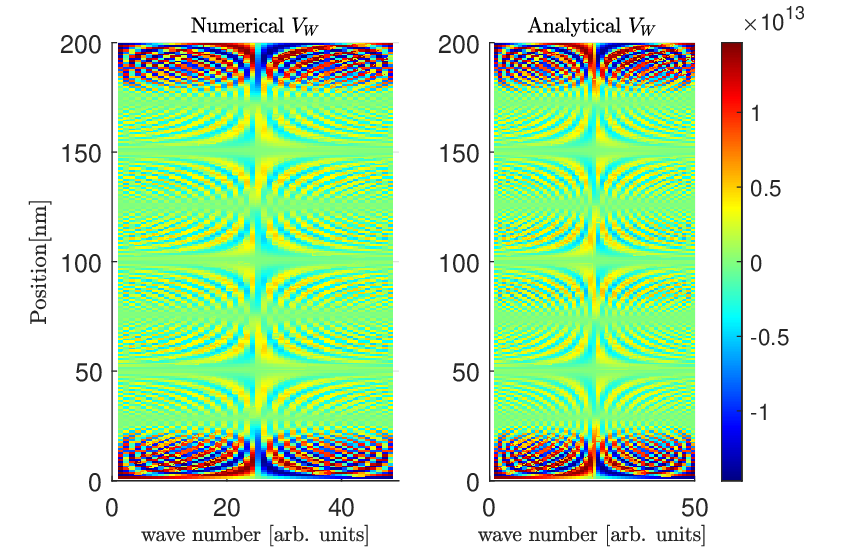}
		\caption{Comparison of actual and estimated Wigner kernel for a quadrupled Gaussian potential function.}
		\label{fig:quad_gaussian_kernel}
	\end{figure}
	\begin{figure}[h!]
		\centering
		\includegraphics[width=1\linewidth]{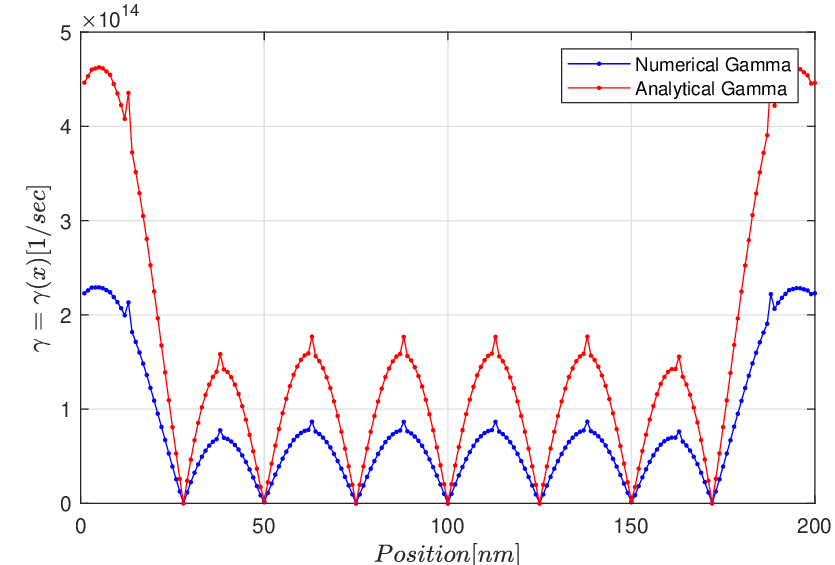}
		\caption{Comparison of Gamma function using the actual and estimated Wigner kernel for a quadrupled Gaussian potential function.}
		\label{fig:quad_gaussian_gamma}
	\end{figure}
	%
	%
	\subsection{Comparison} 
	
	In this section, we compare the results obtained using the proposed analytical method with those derived from the numerical approach by evaluating the execution time of the kernel function, $V_W(x;p)$. To this end, we define $N_x$ as the number of cells in the 
	$x$-direction. As shown in Tables \ref{tab:comparisons_analytical} and \ref{tab:comparisons_numerical}, the average execution times for various setups have been calculated. The results demonstrate that by leveraging the proposed analytical formula in \eqref{eq:last_estimation} significantly reduces the execution time for computing the kernel function compared to the numerical method based on \eqref{eq:wigner_original}. The reported execution times correspond to the average time required to compute $V_W(x;p)$ over 1000 independent runs using the numerical method.
	\begin{table*}[h!]
		\centering
		\begin{tabular}{|c|c|c|c|c|c|}
			\hline
			\diagbox{\# Gaussian}{$N_x$} & 50 & 100 & 150 & 200 & 500 \\ \hline
			Single  & 0.005 & 0.013  & 0.031 & 0.062 & 0.427   \\ \hline
			Double      &  0.003  & 0.014 & 0.033 & 0.061 & 0.433   \\ \hline
			Triple     &   0.003 & 0.014 & 0.034 & 0.067 & 0.428   \\ \hline
			Quadruple   & 0.005 & 0.016  & 0.036 & 0.067 & 0.491  \\ \hline
		\end{tabular}
		\caption{Average execution time for computing $V_W$ (ms) using analytical method in 1000 runs}
		\label{tab:comparisons_analytical}
	\end{table*}
	\begin{table*}[h!]
		\centering
		\begin{tabular}{|c|c|c|c|c|c|}
			\hline
			\diagbox{\# Gaussian}{$N_x$} & 50 & 100 & 150 & 200 & 500 \\ \hline
			Single  & 0.017 & 0.123 & 0.424 & 1.163 & 26.006     \\ \hline
			Double  & 0.016 & 0.120 & 0.430  & 1.127 & 28.016  \\ \hline
			Triple     & 0.019  & 0.127  & 0.436 & 1.161 & 24.825   \\ \hline
			Quadruple  & 0.016  & 0.131  & 0.430 & 1.119 & 27.235  \\ \hline
		\end{tabular}
		\caption{Average execution time for computing $V_W$ (ms) using numerical method in 1000 runs}
		\label{tab:comparisons_numerical}
	\end{table*}
	\section{Conclusion}\label{sec:conclusion}
	
	This work proposed an analytical approximation method for estimating the Wigner kernel and the associated Gamma function within the Monte Carlo Signed Particle Formulation. By approximating the potential-function difference using the leading term of its Maclaurin series expansion, the proposed method avoids costly numerical integration and provides a closed-form expression for efficient kernel evaluation. Numerical experiments on single, double, triple, and quadruple Gaussian potentials showed that the approximated Gamma function closely follows the behavior of the numerically computed one while substantially reducing the execution time.	Future work will focus on generalizing the proposed approach to multi-dimensional phase-space settings, including three-dimensional one-body systems and many-body quantum problems.

	\bigskip
	{\bf{Acknowledgment}. }This work was funded and supported by Ericsson Montr\'eal. J.M. Sellier thanks Maria Anti for her continuous support and enthusiasm.  
\bibliography{cas-refs}
\bibliographystyle{unsrt}

\section*{Biography}
\bio{figs/mehran}
Mehran Attar received the Ph.D. degree in Information and Systems Engineering from Concordia University, Montreal, QC, Canada. He is currently a Postdoctoral Researcher with Ericsson, Montreal, Canada and École de Technologie Supérieure (ÉTS). His research interests include data-driven control and safety of cyber-physical systems, machine learning, cybersecurity, quantum-inspired optimization, and AI-driven methods for wireless communication and control applications.
\endbio
\bio{figs/bassant}
Bassant Selim is currently an Assistant Professor in the Systems Engineering department of the École de Technologie Supérieure (ÉTS). She received her MSc from Université Pierre et Marie, France (2011), and her PhD from Khalifa University, United Arab Emirates (2017). Prior to joining ÉTS, she has worked as a senior data scientist in Ericsson Montreal’s Global Artificial Intelligence Accelerator (GAIA). She also has previous experiences working with Hydro Quebec and Orange Labs (Paris, France). Her research interests include wireless communications, signal processing, communications systems, machine learning, and the internet of things. Bassant has contributed to several industrial patents and more than 30 scientific publications.
\endbio
%
%
\bio{figs/JM}
Jean Michel Sellier has a PhD in Mathematics, 15 years of experience in semiconductor physics, and eight years in the field of machine learning. Today, he is a Data Scientist in GAIA Ericsson, Canada. In the past, he had the honor to create a new formulation of quantum mechanics, which is nowadays known as the signed particle formulation. He is also the creator of a novel quantum-inspired algorithm to train neural networks.
\endbio
\end{document}